\documentclass[]{article}
\usepackage[utf8]{inputenc}
\usepackage{amsmath,amsfonts,color}
\usepackage{graphicx}
\usepackage[affil-it]{authblk}
\usepackage[a4paper,margin=1in]{geometry}
\numberwithin{equation}{section}
\allowdisplaybreaks

\newmuskip\pFqskip
\mathchardef\pFcomma=\mathcode`, 

\newcommand*\pFq[5]{%
  \begingroup
  \begingroup\lccode`~=`,
    \lowercase{\endgroup\def~}{\pFcomma\mkern\pFqskip}%
  \mathcode`,=\string"8000
  {}_{#1}F_{#2}\biggl(\genfrac..{0pt}{}{#3}{#4} \Big| #5\biggr)%
  \endgroup
}

\newcommand*{\Scale}[2][4]{\scalebox{#1}{$#2$}}%

\title{A superintegrable model with reflections on $S^{n-1}$ \\ and the higher rank Bannai-Ito algebra}
\author{Hendrik De Bie}
\affil{Department of Mathematical Analysis, Faculty of Engineering and Architecture, Ghent University, \protect\\Galglaan 2, 9000 Ghent, Belgium \authorcr E-mail address: Hendrik.DeBie@UGent.be\protect\\ \ } 
\author{Vincent X. Genest}
\affil{Department of Mathematics, Massachusetts Institute of Technology, 77 Massachusetts Ave., \protect\\Cambridge, MA 02139, USA \authorcr E-mail address: vxgenest@mit.edu \protect\\ \ }
\author{Jean-Michel Lemay}
\author{Luc Vinet}
\affil{Centre de Recherches Math\'ematiques, Universit\'e de Montr\'eal, C.P. 6128, Succ. Centre-ville,\protect\\ Montr\'eal, QC, Canada, H3C 3J7 \authorcr E-mail address: jean-michel.lemay.1@umontreal.ca, vinet@crm.umontreal.ca}
\date{}

\begin{document}
\maketitle

\begin{abstract}

\noindent A quantum superintegrable model with reflections on the $(n-1)$-sphere is presented. Its symmetry algebra is identified with the higher rank generalization of the Bannai-Ito algebra. It is shown that the Hamiltonian of the system can be constructed from the tensor product of $n$ representations of the superalgebra $\mathfrak{osp}(1|2)$ and that the superintegrability is naturally understood in that setting. The separated solutions are obtained through the Fischer decomposition and a Cauchy-Kovalevskaia extension theorem.
 
\end{abstract}

\section{Introduction}

  This paper introduces a superintegrable model with reflections on the $(n-1)$-sphere which has the rank $(n-2)$-Bannai-Ito algebra as its symmetry algebra. 

  Maximally superintegrable quantum Hamiltonians in $n$ dimensions possess $2n-1$ algebraically independent constants of motion (including $H$). Often interesting in their own right, these systems form the bedrock for the analysis of symmetries and their study has witnessed significant advances in recent years. Of note is the complete classification of all scalar superintegrable systems in two dimensions whose conserved quantities are at most of order two in momenta \cite{2013_Kalnins&Miller&Post_2DContractions,2013_Miller&Post&Winternitz_SI&apps}. It shows that (for Euclidean signature) all superintegrable systems in that class can be obtained as contractions or special cases of the so-called generic model on the $2$-sphere. This developped in parallel with the study of non-linear algebras \cite{Zhedanov91,Zhedanov92} associated to the bispectrality of the orthogonal polynomials of the Askey tableau \cite{Askey_scheme}. These algebras are usually referred to by the name of the corresponding polynomials.

  Separated wavefunctions of the generic model are obtained as joint eigenfunctions of $H$ and one constant of motion. The overlaps between wavefunctions associated to the diagonalization of two different generators are given in terms of Racah polynomials \cite{Miller07}. In view of this, it was somewhat natural to find that the symmetry algebra of the generic scalar system on the $2$-sphere is the Racah algebra \cite{Zhedanov88}. This was subsequently put into a cogent framework when it was observed that the description of the generic model could be formulated via the recoupling of three $\mathfrak{sl}(2)$ realizations \cite{V20131,V20132,V20133}.

  Superalgebras somehow subsumes Lie algebras. Indeed Lie algebras can be engendered as even subalgebra of superalgebras from quadratic expressions in the odd generators of the latter. A translation of this has appeared in the realm of superintegrable models. It has indeed been seen \cite{BIrevue} that by using reflection operators, a number of generic superintegrable models with different constants can be combined together in a supersymmetric fashion to give a generalized Hamiltonian whose symmetries form the Bannai-Ito algebra \cite{BI,BIrevue}. This algebra associated to the Bannai-Ito polynomials \cite{BIbook} has simple defining relations and arises upon considering the tensor product of three $\mathfrak{osp}(1|2)$ superalgebras \cite{BIcoef, laplacedunkl}. It has further been seen that the Racah algebra can be embedded in the Bannai-Ito algebra by using quadratic polynomials in the generators of the latter \cite{embed}. This model with reflection operator together with its symmetry algebra could thus be viewed as more basic than the generic scalar one.

  In the wake of the classification of two-dimensional systems, the exploration of higher dimensional superintegrable models was undertaken. The generic model on the $3$-sphere was shown to be connected to the bivariate Racah polynomials \cite{2013_Kalnins&Miller&Post_2varWilson}; the model with reflections on $S^3$ was also constructed and analyzed \cite{LaplaceDunklS3}. Having in mind the recoupling framework for superintegrable models, it becomes clear that the search for the symmetry algebras of higher dimensional version of the generic model amounts to the identification of the Bannai-Ito and Racah algebras of higher ranks. Results in this direction have been obtained. 

  The higher rank Bannai-Ito algebra was first identified using the Dirac-Dunkl equation as model \cite{higherrankBIalgebra}. It was subsequently constructed in \cite{proc16} using $n$-fold products of $\mathfrak{osp}(1|2)$. Similarly, the Racah algebra was extended to arbitrary ranks in \cite{wouter16} by considering multiple tensor products of $\mathfrak{sl}(2)$ realized in terms of Dunkl operators. It was also observed \cite{iliev16} that the generators of this algebra realize the Drinfeld-Kohno relations.

  We now bring these advances to bear on superintegrable models by providing here the non-relativistic Hamiltonian with reflections that has for symmetry algebra the higher rank Bannai-Ito algebra recently discovered. We shall also show how these symmetries can be put to use in order to obtain the wavefunctions of this quantum model. 

  The paper is divided as follows. In section 2, the model with $n$ parameters on $S^{n-1}$ is introduced and its symmetries are given. In section 3, it is shown how it is built out of $n$ copies of the superalgebra $\mathfrak{osp}(1|2)$. This construction will lead to the identification of the symmetry algebra. The wavefunctions are obtained in Section 4 using the Fischer decomposition and a Cauchy-Kovalevskaia extension theorem. (It will be of interest to observe that such a theorem can also be devised in a scalar situation outside the usual Clifford algebra context.) Some concluding remarks are offered in section 5.

\section{A superintegrable model on $S^{n-1}$}

  Let $s_1,s_2,\dots, s_n$ be the Cartesian coordinates of the $n$-dimensional Euclidean space and consider the embedding of the $S^{n-1}$ sphere given by the constraint: $\sum_{i=1}^n s_i^2=1$. We shall be interested in the system with $n$ parameters $\mu_1, \mu_2, \dots, \mu_n$ with $\mu_i\geq 0$ for $i=1,2,\dots,n$ governed by the Hamiltonian  
  \begin{align} \label{H}
  H = \sum_{1\le i<j\le n} J_{ij}^2 +\sum_{i=1}^n \frac{\mu_i}{s_i^2}(\mu_i - R_i),
  \end{align}
  where 
  \begin{align}
  J_{ij} = \frac{1}{i}(s_i \partial_{s_j} - s_j \partial_{s_i}), \qquad \qquad  R_i f(s_i) = f(-s_i),
  \end{align}
  are the angular momentum operators and reflection operators, respectively. Denote by $[n]$  the set $\{1, 2, 3, ...,n\}$. The quantities
  \begin{align}
  M_A = \Bigg(-\frac{1}{2}+\sum_{i\in A}(\tfrac{1}{2}+\mu_iR_i) + \sum_{\substack{j<k\\ j,k\in A}}\left(-iJ_{jk}-s_k\frac{\mu_j}{s_j}R_j+s_j\frac{\mu_k}{s_k}R_k\right)\prod_{l=j}^{k-1}R_l \Bigg) \prod_{m\in A} R_m,
  \end{align}
  labelled by subsets $A\subset [n]$ can be seen to commute with $H$ and are thus conserved. Note that when the set $A$ contains only one element, say $i$, $M_i$ will be taken to be
  \begin{align}
  M_i = \mu_i, \qquad i=1,2,\dots,n. 
  \end{align}
  In view of their expression, the number of algebraically independent constant of motion $M_A$ will correspond to the number of independent generators of $\mathfrak{so}(n)$ thus implying that $H$ is superintegrable. Furthermore, a direct computation shows that all reflections are also symmetries of $H$:
  \begin{align}
  [H,R_i]=0, \qquad i=1,2,\dots,n.
  \end{align}
  
\section{Algebraic construction from $\mathfrak{osp}(1|2)$}  

  We shall now explain the relation that the Hamiltonian $H$ has with $\mathfrak{osp}(1|2)$. This superalgebra has 5 generators, two odd $x$ and $D$ and three even $E, |x|^2$ and $D^2$ which satisfy the commutation relations:
  \begin{align} \label{osprelns}
  \begin{aligned}
  &\{x,x\}=2|x|^2,  &&\{D,D\}=2D^2, \\
  &\{x,D\}=2E,    &&[D,E]=D, \\
  &[D,|x|^2]=2x,  &&[E,x]=x, \\
  &[D^2,x]=2D,    &&[D^2,E]=2D^2,\\
  &[D^2,|x|^2]=4E, &&[E,|x|^2]=2|x|^2,
  \end{aligned}
  \end{align}
  where $[a,b]=ab-ba$ is the commutator and $\{a,b\}=ab+ba$ is the anti-commutator. One can realize mutually commuting copies of this superalgebra by taking
  \begin{align} \label{irealization}
  \begin{aligned}
  &D_i = \partial_{s_i} - \frac{\mu_i}{s_i}R_i,  &&D^2_i= D_iD_i, \\
  &x_i=s_i, &&|x_i|^2 = s_i^2, \\
  &E_i=s_i\partial_{s_i}+\frac{1}{2},
  \end{aligned}
  \end{align}
  where $i=1,2,\dots,n$. Each superalgebra possesses a sCasimir element given by
  \begin{align}
  S_i = \frac{1}{2}([D_i,x_i]-1),
  \end{align}
  which anticommutes with the odd generators
  \begin{align}
  \{S_i,D_i\}=\{S_i,x_i\}=0,
  \end{align}
  and thus commutes with the even generators
  \begin{align}
  [S_i,E_i]=[S_i,|x_i|^2]=[S_i,D_i^2]=0.
  \end{align}
  It is immediate to verify that in the realization \eqref{irealization}, the reflection $R_i$ obeys the same commutation relations as the sCasimir
  \begin{align}
  [R_i,E_i]=[R_i,|x_i|^2]=[R_i,D_i^2]=\{R_i,D_i\}=\{R_i,x_i\}=0.
  \end{align}
  This implies that one can construct a Casimir operator of the form
  \begin{align}
  Q_i = S_iR_i.
  \end{align}
  It is straightforward to verify that $Q_i$ indeed commutes with every generator. The commuting realizations of $\mathfrak{osp}(1|2)$ can be used as building blocks to construct other realizations. Let $[n]=\{1,2,\dots,n\}$ and $A\subset [n]$ as before. The operators given by
  \begin{align} \label{Arealization}
  \begin{aligned}
  &D_A = \sum_{i\in A}\Big( D_i\prod_{j=1}^{i-1}R_j\Big), &&D^2_A= D_AD_A, \\
  &x_A=\sum_{i\in A}\Big(s_i\prod_{j=1}^{i-1}R_j\Big),    &&|x_A|^2 = \sum_{i\in A}s_i^2, \\
  &E_A=\sum_{i\in A}E_i, &&R_A = \prod_{i\in A} R_i,
  \end{aligned}
  \end{align}
  verify the commutation relations \eqref{osprelns} for any $A\subset [n]$ and thus form new realizations of $\mathfrak{osp}(1|2)$. These result from the repeated application of the coproduct of $\mathfrak{osp}(1|2)$ (see \cite{higherrankBIalgebra}). For any $A$ the sCasimir and the Casimir operators are again defined by
  \begin{align} \label{Arealization2}
  S_A = \frac{1}{2}([D_A,x_A]-1), \quad\qquad Q_A = S_A R_A.
  \end{align}
  One can directly check that
  \begin{align}
  Q_A = M_A, \qquad A\subset[n].
  \end{align}
  Another explicit computation gives
  \begin{align} \label{HsCasimir}
  S_{[n]}^2-S_{[n]}-\frac{(n-1)(n-3)}{4} \ = \sum_{1\le i<j\le n} J_{ij}^2 +\left(\sum_{i=1}^n s_i^2\right)\sum_{i=1}^n \frac{\mu_i}{s_i^2}(\mu_i - R_i).
  \end{align}
  However, since $|x_{[n]}|^2=\sum_{i=1}^n s_i^2$ commutes with $S_{[n]}$ and all the Casimirs, it is central and can be treated as a constant. Taking $|x_{[n]}|^2=1$, it is straightforward to see from \eqref{H} and \eqref{HsCasimir} that
  \begin{align} \label{algebraicH}
  S_{[n]}^2-S_{[n]}-\frac{(n-1)(n-3)}{4} = H.
  \end{align}
  Hence, a quadratic combination of the sCasimir of $n$ copies of $\mathfrak{osp}(1|2)$ yields the Hamiltonian of the superintegrable model presented in section 1 and the $\mathfrak{osp}(1|2)$ Casimirs $Q_A$ will be its symmetries since $[Q_A,H]=0$ for $A \subset [n]$ in view of \eqref{algebraicH}. This makes the Bannai-Ito algebra the symmetry algebra of the model since it is precisely defined as the algebra generated by the $\mathfrak{osp}(1|2)$ intermediate Casimir operators $Q_A$ given by \eqref{Arealization} and \eqref{Arealization2}. The defining relations have been obtained \cite{higherrankBIalgebra} and read: 
  \begin{align} \label{BIrank2}
  \{ Q_A,Q_B\} = Q_{(A\cup B)\setminus (A\cap B)}+2Q_{A\cap B}Q_{A\cup B} + 2Q_{A\setminus (A\cap B)}Q_{B\setminus (A\cap B)},
  \end{align}
  where $A,B\subset[n]$ and $Q_\emptyset = -1/2$. When $n=3$, let $K_1 = Q_{\{1,2\}}, K_2 = Q_{\{2,3\}}$ and $K_3 = Q_{\{1,3\}}$. The recurrence relations \eqref{BIrank2} can then be rewritten as
  \begin{align}
   \{K_1,K_2\}=K_3+\omega_3, \quad \{K_2,K_3\}=K_1+\omega_1, \quad \{K_3,K_1\}=K_2+\omega_2, 
  \end{align}
  where $\omega_1, \omega_2, \omega_3$ are central elements given by
  \begin{align}
   \omega_1 =2Q_3Q_{123}+2Q_1Q_2, \quad \omega_2 = 2Q_1Q_{123}+2Q_2Q_3, \quad \omega_3 = 2Q_2Q_{123}+2Q_1Q_3.
  \end{align}
  This corresponds to the Bannai-Ito algebra already seen \cite{laplacedunkl} to be the symmetry algebra of the $S^2$ version of the Hamiltonian $H$ given in \eqref{H}.

\section{Wavefunctions}
  We shall indicate in this section how the separated wavefunctions of $H$ can be obtained by exploiting the symmetries that have been exhibited. To that end we shall design in this scalar context an extension map of the Cauchy-Kowalevskaia type that is formulated in terms of Dunkl operators. In order to make these operators appear we shall first perform the gauge transformation
  \begin{align}
  z\to \tilde{z}\equiv G(\vec{s})^{-1} z G(\vec{s}), \qquad G(\vec{s})=\prod_{i=1}^n |s_i|^{\mu_i},
  \end{align}
  where $z$ is any operator and $\vec{s}\equiv(s_1,s_2,\dots,s_n)$. Under this transformation, the $\mathfrak{osp}(1|2)$ generators of the realization \eqref{irealization} become
  \begin{align} \label{tildeRealization}
  \begin{aligned}
   &\tilde{D_i} = \partial_{s_i} + \frac{\mu_i}{s_i}(1-R_i),\quad &&\tilde{D}^2_i= \tilde{D}_i\tilde{D}_i,  \\
   &\tilde{x_i} = x_i = s_i, && |\tilde{x}_i|^2=s_i^2,\\
   &\tilde{E_i} = s_i\partial_{s_i} +\gamma_i, &&\tilde{R_i} = R_i,\\
   &\tilde{S_i}=-\mu_iR_i, &&\tilde{Q_i} = \mu_i ,
  \end{aligned}
  \end{align}
  where
  \begin{align}
  \gamma_A = \sum_{i\in A}(\mu_i + \tfrac{1}{2}). 
  \end{align}
  These operators also verify \eqref{osprelns} and correspond to the realization of $\mathfrak{osp}(1|2)$ (or equivalently of $sl_{-1}(2)$) associated to the one-dimensional parabose oscillator \cite{dunklosc}. The construction \eqref{Arealization} can be repeated to obtain operators $\tilde{D}_A, \tilde{x}_A, \tilde{E}_A,\tilde{S}_A$ and $\tilde{Q}_A$ that are gauge equivalent to those without tildes. We can hence obtain the eigenvalues and eigenfunctions of $H$ by finding those of $\tilde{S}_{[n]}$. Note that since $\tilde{S}_{[n]}$ commutes with $P=\prod_{i=1}^n R_i$, this is equivalent to finding the eigenfunctions of $\tilde{Q}_{[n]}$. 

  We now wish to obtain the polynomial eigenfunctions of $\tilde{S}_{[n]}$. Denote by $\mathcal{P}_m(\mathbb{R}^n)$ the space of homogeneous polynomials of degree $m$ in the variables $s_1,s_2,\dots,s_n$ and define $\mathcal{K}_m(\mathbb{R}^n)$ by 
  \begin{align}
   \mathcal{K}_m(\mathbb{R}^n) = \ker \tilde{D}_{[n]} \cap \mathcal{P}_m(\mathbb{R}^n)
  \end{align}
  with $\ker \tilde{D}_{[n]}$ the set of null-eigenfunctions of $\tilde{D}_{[n]}$. $\mathcal{K}_m(\mathbb{R}^n)$ is an eigenspace of $\tilde{S}_{[n]}$. Indeed, take $\psi_m \in \mathcal{K}_m(\mathbb{R}^n)$, one has
  \begin{align*}
   \tilde{S}_{[n]}\tilde{\psi}_m &= \tfrac{1}{2}(\tilde{D}_{[n]}\tilde{x}_{[n]}-\tilde{x}_{[n]}\tilde{D}_{[n]}-1)\tilde{\psi}_m = \tfrac{1}{2}(\tilde{D}_{[n]}\tilde{x}_{[n]}-1)\tilde{\psi}_m \\
    &=\tfrac{1}{2}(\tilde{D}_{[n]}\tilde{x}_{[n]}+\tilde{x}_{[n]}\tilde{D}_{[n]}-1)\tilde{\psi}_m = \tfrac{1}{2}(\{\tilde{x}_{[n]},\tilde{D}_{[n]}\}-1)\tilde{\psi}_m \\
    &=\tfrac{1}{2}(2\tilde{E}_{[n]}-1)\tilde{\psi}_m = \left[\sum_{i=1}^n s_i\partial_{s_i} +\gamma_{[n]}-\tfrac{1}{2}\right]\tilde{\psi}_m,
  \end{align*}
  where we used the property $\tilde{D}_{[n]}\tilde{\psi}_m=0$ and the commutation relations \eqref{osprelns}. Since $\tilde{\psi}_m$ is a homogeneous polynomial of degree $m$, it is an eigenfunction of the Euler operator : $\sum_{i=1}^n s_i\partial_{s_i} \tilde{\psi}_m = m \tilde{\psi}_m $. This implies that 
  \begin{align} \label{eigenv}
   \tilde{S}_{[n]}\tilde{\psi}_m = (m+\gamma_{[n]}-\tfrac{1}{2})\tilde{\psi}_m
  \end{align}
   and concludes our proof. 

  Our aim is thus to construct a basis for $\mathcal{K}_m(\mathbb{R}^n)$. This will be done by relying on two constructs. One is a Cauchy-Kovalevskaia (CK)-map between the space of homogeneous polynomials of degree $m$ in $n-1$ variables $\mathcal{P}_m(\mathbb{R}^{n-1})$ and the space $\mathcal{K}_m(\mathbb{R}^n)$ of null-eigenfunctions of $\tilde{D}_{[n]}$ that are homogeneous polynomials of degree $m$ in $n$ variables:
   \begin{align} \label{CKmap}
    {\bf CK}_{s_n}^{\mu_n} : \mathcal{P}_m(\mathbb{R}^{n-1}) \to \mathcal{K}_{m}(\mathbb{R}^n).
   \end{align}
   To construct explicitly the map ${\bf CK}_{s_n}^{\mu_n}$ take $p(s_1,\dots,s_{n-1})\in \mathcal{P}_m(\mathbb{R}^{n-1})$ and let
   \begin{align}
   {\bf CK}_{s_n}^{\mu_n}[p(s_1,\dots,s_{n-1})]=\sum_{\alpha=0}^{m}s_n^\alpha p_\alpha(s_1,\dots,s_{n-1}),
  \end{align}
  where $p_\alpha(s_1,\dots,s_{n-1})\in \mathcal{P}_{m-\alpha}(\mathbb{R}^{n-1})$ and $p_0(s_1,\dots,s_{n-1})\equiv p(s_1,\dots,s_{n-1})$. Demand that 
  \begin{align}
  \tilde{D}_{[n]}\sum_{\alpha=0}^{m}s_n^\alpha p_\alpha(s_1,\dots,s_{n-1})=0 
  \end{align}
  and solve for the coefficients $p_\alpha(s_1,\dots,s_{n-1})$. A straightforward calculation yields
  \begin{align}
   {\bf CK}_{s_n}^{\mu_n}= \sum_{i=0}^{\infty} \frac{(-1)^i(s_n)^{2i}}{i!(\gamma_n)_i(2)^{2i}}\tilde{D}_{[n-1]}^{2i} + \sum_{i=0}^{\infty} \frac{(-1)^{i+1}(s_n)^{2i+1}}{i!(\gamma_n)_{i+1}(2)^{2i+1}}R_{[n-1]}\tilde{D}_{[n-1]}^{2i+1},
  \end{align}
  where $(a)_i = a(a+1)\dots(a+i-1)$ denotes the Pochhammer symbol. It can be shown that the resulting map is an isomorphism (the proof follows the one given in \cite{wouter16}).

  The second construct is the Fisher decomposition which states that the space of homogeneous polynomials $\mathcal{P}_m(\mathbb{R}^{n})$ can be decomposed over spaces $\mathcal{K}_{l}(\mathbb{R}^n)$ as follows :
   \begin{align} \label{Fischer}
    \mathcal{P}_m(\mathbb{R}^n) = \bigoplus_{j=0}^{m} \tilde{x}_{[n]}^j \mathcal{K}_{m-j}(\mathbb{R}^n).
   \end{align}
  (This is analogous to the decomposition of $\mathcal{P}_m(\mathbb{R}^{n})$ over spaces of spherical harmonics.) Upon using in alternance the Fisher decomposition and the CK map, one shows that the space $\mathcal{K}_{m}(\mathbb{R}^n)$ can be represented as follows :
  \begin{align}
    \mathcal{K}_m(\mathbb{R}^n) \cong {\bf CK}_{s_n}^{\mu_n}\Big[\bigoplus_{j_{n-2}=0}^{m} \tilde{x}_{[n-1]}^{m-j_{n-2}}{\bf CK}_{s_{n-1}}^{\mu_{n-1}}\Big[ \dots \bigoplus_{j_1=0}^{j_2} \tilde{x}_{[2]}^{j_2-j_1}{\bf CK}_{s_2}^{\mu_2}\left[\mathcal{P}_{j_1}(\mathbb{R})\right]\Big]\Big].
   \end{align}
  This implies that a basis for $\mathcal{K}_m(\mathbb{R}^n)$ is provided by the eigenfunctions $\{\tilde{\psi}_{j_1,\dots,j_{n-1}}^{(m)}(\vec{s})\}_{\sum_{i=1}^{n-1}j_i=m}$ given by
   \begin{align} \label{cbase}
    \tilde{\psi}_{j_1,\dots,j_{n-1}}^{(m)}(\vec{s}) = {\bf CK}_{s_n}^{\mu_n}\big[ \tilde{x}_{[n-1]}^{j_{n-1}}{\bf CK}_{s_{n-1}}^{\mu_{n-1}}\big[\dots \tilde{x}_{[2]}^{j_2}{\bf CK}_{s_2}^{\mu_2}[s_1^{j_1}]\big]\big].
   \end{align}
  Different bases are obtained by permuting the order in which the CK-extensions are applied. This leads to explicit formulas. The calculation that \eqref{cbase} entails can be carried out straightforwardly with the help of the identities with $\tilde{\psi}_m \in \mathcal{K}_m(\mathbb{R}^n)$ 
   \begin{align}
   \begin{aligned}
   \tilde{D}_{[n]}^{2\alpha}\tilde{x}_{[n]}^{2\beta}\tilde{\psi}_m &= 2^{2\alpha}(-\beta)_\alpha(1-m-\beta-\gamma_{[n]})_\alpha x_{[n]}^{2\beta-2\alpha}\tilde{\psi}_m, \\
   \tilde{D}_{[n]}^{2\alpha}\tilde{x}_{[n]}^{2\beta+1}\tilde{\psi}_m &= 2^{2\alpha}(-\beta)_\alpha(-m-\beta-\gamma_{[n]})_\alpha x_{[n]}^{2\beta+1-2\alpha}\tilde{\psi}_m, \\   
   R_{[n]}\tilde{D}_{[n]}^{2\alpha+1}\tilde{x}_{[n]}^{2\beta}\tilde{\psi}_m &= -2^{2\alpha}\beta(1-\beta)_\alpha(1-m-\beta-\gamma_{[n]})_\alpha x_{[n]}^{2\beta-2\alpha-1}R_{[n]} \tilde{\psi}_m, \\
   R_{[n]}\tilde{D}_{[n]}^{2\alpha+1}\tilde{x}_{[n]}^{2\beta+1}\tilde{\psi}_m &= 2^{2\alpha+1}(-\beta)_\alpha(m+\beta+\gamma_{[n]})(1-m-\beta-\gamma_{[n]})_\alpha x_{[n]}^{2\beta-2\alpha} R_{[n]}\tilde{\psi}, 
    \end{aligned}
   \end{align}
   which follows from \eqref{osprelns}. The results can be presented in terms of the Jacobi polynomials $P_n^{(\alpha,\beta)}(x)$ defined as \cite{Askey_scheme}
   \begin{align}
    P_n^{(\alpha,\beta)}(x) = \frac{(\alpha+1)_n}{n!} \pFq{2}{1}{-n,n+\alpha+\beta+1}{\alpha+1}{\frac{1-x}{2}}
   \end{align}
   and with the help of the identity :
   \begin{align}
   (x+y)^n P_n^{(\alpha,\beta)}\left(\frac{x-y}{x+y}\right) = \frac{(\alpha+1)_n}{n!}x^n\pFq{2}{1}{-n,-n-\beta}{\alpha+1}{-\frac{y}{x}}. 
   \end{align}
   One obtains
   \begin{align}
    \tilde{\psi}_{j_1,\dots,j_{n-1}}^{(m)}(\vec{s}) = {\bf P}_{n}{\bf P}_{n-1}\dots{\bf P}_3 Q_{j_1}(s_1,s_2),
   \end{align}
   where ${\bf P}_{k}$ is an operator depending on the $k$ variables $s_1,\dots,s_k$ and on the $k-1$ parameters $j_1,\dots,j_{k-1}$ which is given by  
   \begin{align*}
   \begin{aligned}
    &{\bf P}_{k} = \frac{c!}{(\gamma_k)_c}\left(\sum_{i=1}^k s_i^2\right)^c \\ 
    &\times \begin{cases}
    \begin{aligned}
    P_c^{(\gamma_k-1,j_{[k-2]}+\gamma_{[k-1]}-1)}\Scale[0.9]{\left(\frac{s_1^2+\dots+s_{k-1}^2-s_k^2}{s_1^2+\dots+s_{k-1}^2+s_k^2}\right)}
    + \Scale[0.9]{\frac{s_k}{s_1^2+\dots+s_k^2}} P_{c-1}^{(\gamma_k,j_{[k-2]}+\gamma_{[k-1]})}\Scale[0.9]{\left(\frac{s_1^2+\dots+s_{k-1}^2-s_k^2}{s_1^2+\dots+s_{k-1}^2+s_k^2}\right)}\tilde{x}_{[k-1]}R_{[k-1]}
    \end{aligned}
     &\Scale[0.9]{\text{if $j_{k-1}=2c$,}} \\[1em]
    \begin{aligned}
    P_c^{(\gamma_k-1,j_{[k-2]}+\gamma_{[k-1]})}\Scale[0.9]{\left(\frac{s_1^2+\dots+s_{k-1}^2-s_k^2}{s_1^2+\dots+s_{k-1}^2+s_k^2}\right)}\tilde{x}_{[k-1]} 
    - \Scale[0.88]{\frac{j_{[k-2]}+c+\gamma_{[k-1]}}{c+\gamma_k}} P_c^{(\gamma_4,j_{[k-2]}+\gamma_{[k-1]}-1)}\Scale[0.9]{\left(\frac{s_1^2+\dots+s_{k-1}^2-s_k^2}{s_1^2+\dots+s_{k-1}^2+s_k^2}\right)}s_{k}R_{[k-1]}
    \end{aligned}
     &\Scale[0.88]{\text{if $j_{k-1}=2c+1$}} 
    \end{cases}
   \end{aligned}
   \end{align*}
   with 
   \begin{align}
    j_{[m]} = \sum_{i=1}^m j_i
   \end{align}
   and where $Q_{j_1}(s_1,s_2)$ is the function
   \begin{align*}
   \begin{aligned}
    Q_{j_1}(s_1,s_2) = \frac{a!}{(\gamma_2)_a}(s_1^2+s_2^2)^a 
    \times \begin{cases}
    \begin{aligned}
    P_a^{(\gamma_2-1,\gamma_1-1)}\Scale[1]{\left(\frac{s_1^2-s_2^2}{s_1^2+s_2^2}\right)}
    + \Scale[1]{\frac{s_1s_2}{s_1^2+s_2^2}} P_{a-1}^{(\gamma_2,\gamma_1)}\Scale[1]{\left(\frac{s_1^2-s_2^2}{s_1^2+s_2^2}\right)}
    \end{aligned}
     &\Scale[1]{\text{if $j_1=2a$,}} \\[1em]
    \begin{aligned}
    s_1 P_a^{(\gamma_2-1,\gamma_1)}\Scale[1]{\left(\frac{s_1^2-s_2^2}{s_1^2+s_2^2}\right)}
    - \Scale[1]{s_2 \frac{a+\gamma_1}{a+\gamma_2}} P_a^{(\gamma_2,\gamma_1-1)}\Scale[1]{\left(\frac{s_1^2-s_2^2}{s_1^2+s_2^2}\right)}
    \end{aligned}
     &\Scale[1]{\text{if $j_1=2a+1$.}} 
    \end{cases}
   \end{aligned}
   \end{align*} 
  Note that the expressions for ${\bf P}_{k}$ contain the operators $\tilde{x}_{[k-1]}$ and $R_{[k-1]}$ respectively. Recalling the expressions \eqref{tildeRealization} and \eqref{Arealization}, it can be seen that these operators only involve the variables $s_i$ and the reflection operators $R_i$ with $i=1,\dots,k-1$. These reflections conveniently account for signs occuring in the solutions and prevent the need to give different expressions for each parity combination of the parameters $j_1,j_2,\dots,j_{n-1}$.   

  By effecting the reverse gauge transformation, we thus obtain a basis for the eigenspace of the operator $S_{[n]}$ given by
  \begin{align}
   \psi_{j_1,\dots,j_{n-1}}^{(m)}(\vec{s}) = \tilde{\psi}_{j_1,\dots,j_{n-1}}^{(m)}(\vec{s})G(\vec{s}), 
  \end{align}
  where $m=0,1,\dots$ and $j_1+\dots+j_{n-1}=m$. Given \eqref{eigenv}, they obey the relation
  \begin{align}
   S_{[n]}\psi_{j_1,\dots,j_{n-1}}^{(m)}(\vec{s}) = (m+\gamma_{[n]}-\tfrac{1}{2})\psi_{j_1,\dots,j_{n-1}}^{(m)}(\vec{s}).
  \end{align}
  Recalling \eqref{algebraicH}, this is seen to imply that
  \begin{align}
   H\psi_{j_1,\dots,j_{n-1}}^{(m)}(\vec{s})=(m+\gamma_{[n]})(m+\gamma_{[n]}-2)\psi_{j_1,\dots,j_{n-1}}^{(m)}(\vec{s}).
  \end{align}
  The normalized eigenfunctions are given by
  \begin{align}
  \Psi_{j_1,\dots,j_{n-1}}^{(m)}(\vec{s}) = \frac{\nu_1}{\sqrt{2}}\left(\prod_{i=3}^n\eta_i\right)\psi_{j_1,\dots,j_{n-1}}^{(m)}(\vec{s}),
  \end{align}
  where
  \begin{align}
   &\nu_1 = (\gamma_2)_a \sqrt{\frac{\Gamma(a+\gamma_{[2]})}{a!\Gamma(a+\gamma_1)\Gamma(a+\gamma_2)}}\times
            \begin{cases}
            1 \quad &\text{if $j_1 = 2a$,} \\
            \sqrt{\frac{a+\gamma_2}{a+\gamma_1}} \quad &\text{if $j_1=2a+1$,}
            \end{cases} \\           
   &\eta_k = (\gamma_k)_c \sqrt{\frac{\Gamma(c+j_{[n-2]}+\gamma_{[n]})}{c!\Gamma(c+\gamma_n)\Gamma(c+j_{[n-2]}+\gamma_{[n-1]})}}\times
            \begin{cases}
            1 \quad &\text{if $j_{n-1} = 2c$,} \\
            \sqrt{\frac{c+\gamma_n}{c+j_{[n-2]}+\gamma_{[n-1]}}} \quad &\text{if $j_{n-1}=2c+1$,}
            \end{cases}           
  \end{align}
  so that 
  \begin{align}
   \int_{S^{n-1}} \Psi_{j_1,\dots,j_{n-1}}^{(m)\dag}(\vec{s}) \Psi_{k_1,\dots,k_{n-1}}^{(m')}(\vec{s}) d\vec{s} = \delta_{m,m'}\delta_{j_1,k_1}\dots\delta_{j_{n-1},k_{n-1}}.
  \end{align}
  This can be verified directly from the orthogonality relation of the Jacobi polynomials \cite{Askey_scheme}.
\section{Conclusion}
To sum up, we have introduced a new quantum superintegrable model with reflections on the $(n-1)$-sphere. Its symmetries were given explicitly and shown to realize the higher rank generalization of the Bannai-Ito algebra. It was observed that the model can be constructed through the combination of $n$ independent realizations of the superalgebra $\mathfrak{osp}(1|2)$. A quadratic expression in the total sCasimir operator was found to coincide with the Hamiltonian while the intermediate Casimir operators were seen to form its symmetries. The exact solutions have been obtained by using a Cauchy-Kovalevskaia extension theorem.

In keeping with the 2-dimensional picture \cite{BIrevue} the overlap between wavefunctions associated to different maximal Abelian subalgebras \cite{higherrankBIalgebra} of the Bannai-Ito algebra will be expressed in terms of multivariate Bannai-Ito polynomials that we plan to characterize in the near future. 

We have stressed that the Hamiltonian $H$ with reflections on $S^{n-1}$ actually commutes with all reflection operators and that these can hence be diagonalized simultaneously with $H$. In each of the sectors with definite parity, $H$ reduces to a scalar Hamiltonian that extends to $S^{n-1}$ the generic model on $S^2$ known to have the Racah algebra as symmetry algebra. It was shown in \cite{wouter16} that these scalar models on $S^{n-1}$ admit the (more involved) higher rank Racah algebra identified in the same article as the algebra generated by intermediate Casimir operators in the $n$-fold tensor product of realizations of $\mathfrak{sl}(2)$. This indicates that, as in the rank 1 case \cite{embed}, there is an embedding of the Racah algebra in the Bannai-Ito one for higher ranks also. This rests on the fact that the $\mathfrak{sl}(2)$ (intermediate) Casimir operators are quadratic expressions that are reflection invariant in the (intermediate) Casimir operators of $\mathfrak{osp}(1|2)$. Details will be given elsewhere. 

While in two dimensions, all scalar second order superintegrable models (with Euclidean signature) can be obtained from the generic model, this is obviously not so in higher dimensions. We have here provided a superintegrable multidimensional version with reflections of the master model in two dimensions; There are however other known superintegrable models in arbitrary dimensions that do not derive from the one discussed here. Of particuliar interest are the rational Calogero models which are formulated in terms of Dunkl and reflection operators especially when distinguishable particles are considered. These models are superintegrable and their symmetries have been much studied. (See for instance \cite{Kuznetsov,Feigin} among the many references on this topic.) We intend to revisit the rational Calogero model with the perspective on superintegrable systems brought in this paper. 

\section*{Acknowledgments}
The research of HDB is supported by the Fund for Scientific Research-Flanders (FWO-V), project ``Construction of algebra realisations using Dirac-operators'', grant G.0116.13N. VXG holds a postdoctoral fellowship from the Natural Science and Engineering Research Council of Canada (NSERC). JML holds an Alexander-Graham-Bell fellowship from
NSERC. The research of LV is supported in part by NSERC.



\end{document}